# Evaluating Proposed Spatiotemporal Matching for Scope 2 Market-based Accounting: emissions, cost, and policy tradeoffs for hourly matching and tighter geographic constraints

Authors: Daniel Livengood[1], Arin Kaye[1]*, Nils Johnson[1], John Bistline[2], Aranya Venkatesh[1], Bailie Neary[1], Adam Diamant[1]

**Affiliations:**

[1] Electric Power Research Institute; Palo Alto, USA.

[2] Watershed Technology, Inc.; San Francisco, USA.

*Corresponding author. Email: AKaye@epri.com

**Abstract**

Corporate voluntary electricity procurement represents a large share of clean energy investment, yet accounting rules for the electricity and associated emissions claims are contested. Whether tighter spatiotemporal constraints on carbon-free electricity would substantially reduce emissions or primarily raise compliance costs is unknown. Using an integrated capacity expansion model with flexible alternate scenarios of deliverability zones and policy contexts, this analysis suggests that across most scenarios, geographically narrower deliverability zones are associated with greater consequential emissions reductions along with higher EAC prices. National results show that in the Reference Technology scenario, the model preferentially deploys natural gas with CCS to meet new CFE demand. In the Limited Technology scenarios where natural gas with CCS is unavailable, wind and solar resources are deployed. Individual regions show varying costs of $CO_2$ abatement under the scenarios, with the highest EAC prices occurring under Limited Technology assumptions and Fine spatial boundaries. This regional cost variation for CFE procurement could have implications for future load siting decisions or for which existing loads CFE will be procured to match. These findings highlight the importance of clearly defining "deliverability zones" in shaping both the emissions impact and cost of voluntary CFE procurement.

## 1. Introduction

Corporate carbon-free electricity (CFE) procurement has become a central mechanism used by companies to reduce their reported scope 2 greenhouse gas (GHG) emissions associated with buying and using electricity in their operations. Voluntary clean electricity purchases via power purchase agreements (PPAs), green tariffs, and other instruments account for a growing proportion of new clean energy investments globally. The accounting frameworks that apply to such transactions influence the degree to which voluntary efforts reduce system-level emissions. Emissions reductions associated with voluntary procurement depend on how energy attribute certificates (EACs), which are market instruments used to track clean electricity claims, are defined and matched to actual end-use electricity consumption. These EACs carry an emission factor (EF) reflecting the operational emissions intensity of the underlying generation of that MWh (e.g., solar generation has a near-zero [~0 $tCO_2$/MWh] operational emission factor). However, the standards that influence EAC usage in scope 2 market-based accounting are contested, and proposed revisions to existing standards could alter the geography and cost of participating in voluntary clean energy markets.

The current framework for corporate emissions accounting for purchased electricity is based on the Greenhouse Gas Protocol (GHGP) Scope 2 Guidance. The scope 2 market-based accounting method allows companies to substitute emission factors associated with contracted electricity resources for the emissions intensities of grid-based electricity that is dispatched and used by these entities, provided they hold qualifying EACs. Under this market-based method (MBM), EACs can be purchased anywhere in the country and matched to annual consumption on a volumetric basis [1]. The MBM guidance allows reporting entities to report their MBM emissions based on the EF associated with contracted energy resources and EACs, rather than the EF associated with the physical delivery of electricity to the end-use load. The use of the MBM is facilitated by the issuance, trade, tracking, and retirement of EACs, which can be bundled and sold along with the underlying electricity or unbundled and purchased separately from the underlying electricity that created the EAC. Critics have argued that this flexibility allows companies to claim emission factors from generation resources that could not have physically supplied their load (e.g., either because the time of consumption does not match the time of electricity production and/or the generation source is not physically tied to the claiming consumer), which misrepresents actual physical emissions [2].

In response, the GHGP's 2025 Scope 2 Consultation [3] proposes an important change to the definition of scope 2 emissions that would clarify that *"…scope 2 must only include emissions from electricity generation processes that are physically connected to the reporter's value chain*." To implement this change, the GHGP proposes two new spatiotemporal requirements to demonstrate the physical connection between electricity supply and consumption when companies use the MBM:

(i) Hourly matching of qualifying generation to end-use load; and

(ii) Geographic "deliverability" constraints that would require qualifying resources to be generated in the same geographic location where the electricity is consumed.

Previous analytical studies, using different approaches, have focused on comparing the impacts of the temporal boundary used for the MBM, showing that the volumetric (i.e., annual) matching of generation to load permitted in the 2015 Guidance can result in substantially fewer $CO_2$ reductions than reported using the MBM or as compared to the hourly matching proposed in the Scope 2 Consultation [4] [5] [6] [7] [8]. Some studies have leveraged historic marginal emission factors to estimate the impacts of CFE procurement (e.g., [7]). Others have used electric sector capacity expansion planning models to estimate the impacts of various GHGP design configurations (for example, in the western U.S. [6] and for some regions in Europe [5]). One study examined the joint effects of temporal and geographic deliverability design choices in the U.S., under two alternate energy policy environments, while another used short-term electric sector operations models to assess how transmission congestion affects carbon accounting [9] [10]. As noted in [11], some of these issues also have been addressed in the literature related to the Inflation Reduction Act's §45V tax credit for electrolytic hydrogen production, although insights may not be directly transferrable to the discussion around the GHGP design considerations, in part due to spatiotemporal differences in load profiles for electrolytic hydrogen and §45V design.

A recent study added to this literature with regional analysis of voluntary CFE demand in the U.S. across several scenarios that varied GHGP program design, including spatial, temporal, and incrementality requirements, in conjunction with technology and policy sensitivities. This study used an integrated energy-economy modeling framework to assess system-wide costs, generation, and emissions impacts [11]. However, the granularity of deliverability zone definitions, ranging from grid-level regions to interconnections to national boundaries, and how that granularity interacts with technology availability and the current U.S. federal and state policies have not been systematically examined. Restricting the geography associated with using qualifying resources in the MBM could have substantial impacts on the voluntary procurement of clean energy resources; data from the EPA Green Power Partnership suggest that virtual PPAs (PPAs that provide owners with unbundled EACs) account for approximately 60% of voluntary renewable PPA sales in the U.S., and corporate buyers are increasingly contracting for low-emission electricity and CFE via PPAs [12].

The analysis in this report builds on [11] to assess the potential system-wide effects of voluntary CFE procurement based on the proposed changes to the scope 2 MBM spatiotemporal requirements with emphasis on evaluating the effects of varying deliverability zone size, from narrow "grid zones" to the national level in the U.S. The study analyzes these impacts against a Counterfactual Reference scenario that includes implementation of all existing federal laws and regulations including the 2025 budget reconciliation bill, the One Big Beautiful Bill Act (OBBBA), and all current state clean energy policies and emissions targets. Additional uncertainties around eligible electric power generation technologies also are considered as sensitivities in this analysis.

The remainder of the paper is structured to foreground the policy relevance of the modeling results. Section 2 describes the model, policy baseline, deliverability zones, and technology sensitivities. Section 3 reports national, regional, price, emissions, and technology-sensitivity results. Section 4 interprets these findings as policy design tradeoffs for Scope 2 market-based accounting, voluntary CFE procurement, and future EAC market design. Section 5 identifies limitations and future research needs, including demand elasticity, finer transmission constraints, and implementation pathways for hourly and deliverability-based accounting.

## 2. Methods and Scenario Design

This analysis uses EPRI's U.S. Regional Economy Greenhouse Gas and Energy (US-REGEN) model to assess system impacts of voluntary CFE procurement under the proposed changes to requirements described in the GHGP's Scope 2 Consultation, with a focus on analysis of potential impacts associated with different geographic deliverability zone requirements. US-REGEN is an integrated energy systems model that includes detailed electric sector capacity expansion and dispatch decisions, along with a representation of fuel supply pathways and technology choice in end-use sectors [13]. The electric and fuels supply model determines the least-cost mix of resources given assumptions about technology costs, markets, and policies while capturing temporal detail between load, wind output, and solar output as well as chronological operations to characterize energy storage and other balancing resources [14] [15]. The US-REGEN end-use model provides hourly estimates of regional load over time based on detailed modeling of technology adoption and utilization across buildings, transport, and industrial sectors [16]. More details about US-REGEN can be found in publicly available documentation [13].

Scenarios in this study vary CFE program design, participation rates (i.e., market size), and assumptions around qualifying generation technologies, as summarized in Table 1. Across all scenarios, CFE targets are applied as a share of commercial and industrial (C&I) load to all model regions simultaneously (see Figure S2 for the categories of total national demand and Figure S3 for the incremental CFE procurement amounts included in the model). The analysis assumes existing CFE demand is matched to existing CFE supply consistent with the provision in the GHGP's Scope 2 Consultation that would permit reporters using the MBM to count legacy contracts and their pro-rata share of qualifying resources that are available to them via Standard Supply Service (SSS). Because of the inclusion of this provision, this study considers only new, incremental voluntary CFE procurement, which is set for modeling purposes at 10% of annual C&I load in each region starting in 2030 which aligns with the growth of voluntary renewable power sales seen over the last decade [12]. Although participation is implemented regionally as a share of C&I load for comparability in these scenarios, actual regional allocations of participation will vary, in part based on costs, policy environment, and other drivers, such as new loads from data centers, which are projected to be greater than 10% of C&I load in 2030 and beyond [17].

In the scenarios that include CFE procurement, four spatial aggregations and two temporal aggregations are analyzed. Scenarios reflecting current GHGP design requirements assume procurement at the U.S. level (where generation anywhere in the country may qualify) with annual volumetric matching. All other CFE procurement scenarios employ hourly matching of qualifying generation and end-use consumption. Intra-annual variation is represented by 120 representative hours that preserve hourly chronology in load profiles generated endogenously by the US-REGEN end-use model, and CFE requirements are met across these representative hours. Under the assumption that all existing CFE procurement in the Reference and other Technology scenarios is met with *existing* CFE resources provided under SSS and the legacy contract clause, the procurement of additional qualifying resources to meet the assumed increase in CFE demand in this study is assumed to be met by new, incremental qualifying power generation resources. For comparison, a Counterfactual Reference scenario considers no voluntary CFE procurement while maintaining the same total demand.

Three levels of regional aggregation are considered in addition to the national level, as shown in Figure 1. Regional aggregation ranges from the most geographically narrow with CFE matching limited to within grid zones (i.e., Fine zone) to groups of grid regions (i.e., Broad zone) to interconnections (i.e., Interconnect). For each regional aggregation scenario, qualifying CFE resources are required to be generated in the same geographic region as the CFE demand.

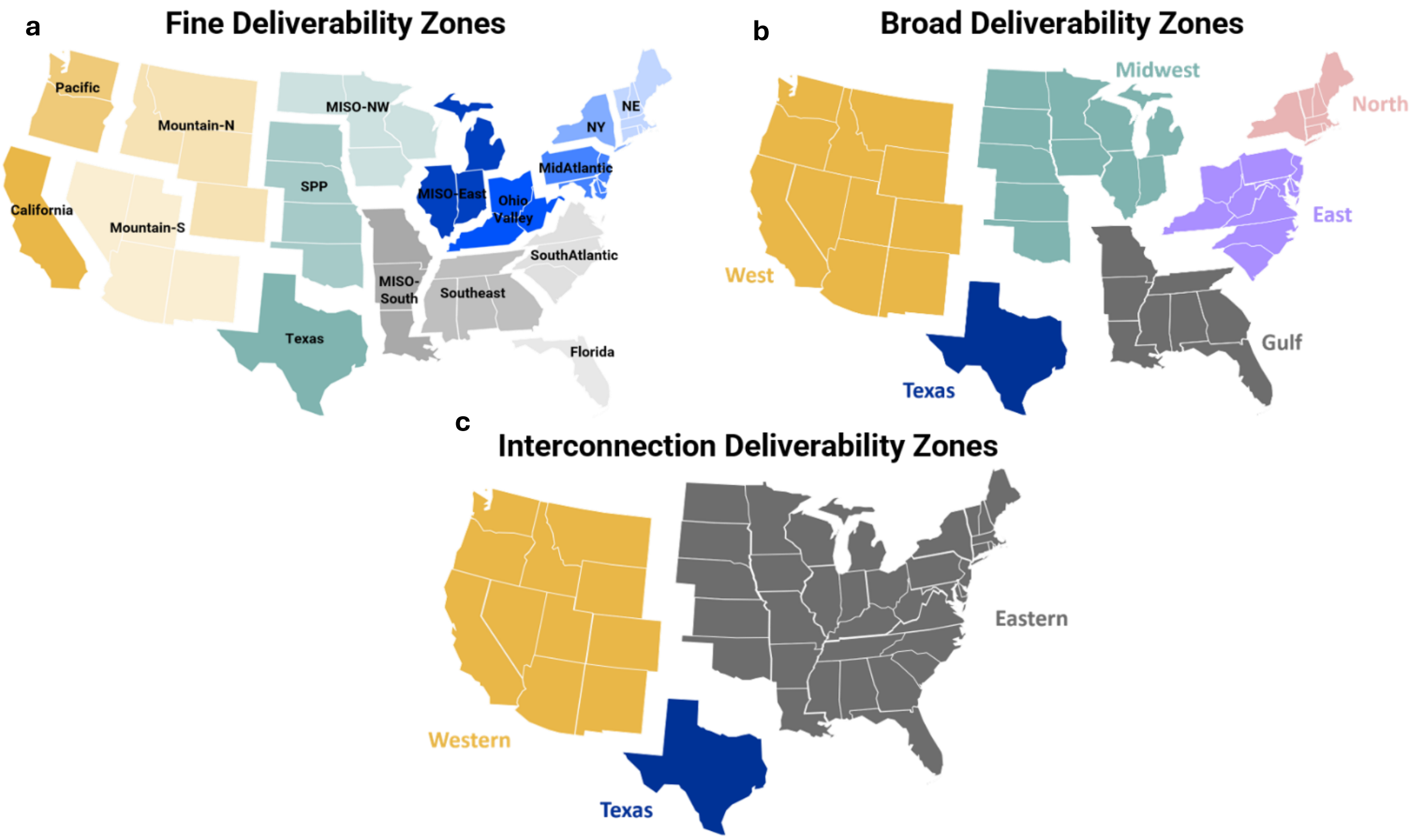

*Figure 1. Sub-national deliverability zones modeled in this study.*

Further, three technology sensitivities are considered:

- Reference Technology: Based on reference capital cost assumptions for wind, solar, storage, and natural gas generation (shown in Figure S1), reference assumptions for biomass availability, and reference assumptions for carbon capture storage costs, performance, and availability (supply curves for CCS in US-REGEN are available in figures 15-17 in [18]).

- Limited Technology: Based on higher-than-reference capital cost assumptions for wind, solar, storage, and natural gas generation, limited biomass availability, and no availability of carbon capture and sequestration (e.g., representing challenges with possible future technology readiness or policy landscape).

- Optimistic Technology: Based on lower-than-reference capital cost assumptions for wind, solar, and storage, reference capital cost assumptions for natural gas generation, reference assumptions for biomass availability, and reference assumptions for carbon capture storage costs, performance, and availability.

This study notes the uncertainty around how and to what extent resources using carbon capture and sequestration (CCS) may be counted as non-emitting generation, leading to an exploration of a range of results both with and without CCS resources qualifying as CFE[1]. As mentioned, the

[1] While the MBM requires entities to report emissions associated with all types of generation resources acquired using power purchase agreements and similar contracts, this analysis focuses on incremental demand for carbon-free electricity resources. Carbon free electricity (CFE) resources in this study refer to renewable energy resources (e.g., wind, solar, geothermal), non-emitting generation resources (e.g., nuclear), and low-emitting resources (e.g., biomass, coal, and natural gas-fired generation paired with carbon, capture and sequestration).

Limited Technology scenarios consider the situation where CCS resources are not available, either due to technology readiness reasons, policy, or other restrictions.

| **Scenario** | **Deliver-ability Zone** | **Temporal matching** | **Wind, solar, and storage capital costs** | **Bioenergy Availability** | **Natural gas capital costs** | **CCS costs** |
|---|---|---|---|---|---|---|
| **Technology:** Reference/Limited/Optimistic **CFE Demand**: No New CFE[2] | N/A | N/A | Reference/ High/ Low | Full/ Limited/ Full | Reference/ High/ Reference | Reference/ Unavailable/ Low |
| **Technology:** Reference/Limited/Optimistic **CFE Zone:** GHGP | 1 U.S. region | Annual | Reference/ High/ Low | Full/ Limited/ Full | Reference/ High/ Reference | Reference/ Unavailable/ Low |
| **Technology:** Reference/Limited/Optimistic **CFE Zones:** Interconnect | 3 Inter-connection zones | Hourly via 120 representative hours | Reference/ High/ Low | Full/ Limited/ Full | Reference/ High/ Reference | Reference/ Unavailable/ Low |
| **Technology:** Reference/Limited/Optimistic **CFE Zones:** Broad | 6 ISO-level zones | Hourly via 120 representative hours | Reference/ High/ Low | Full/ Limited/ Full | Reference/ High/ Reference | Reference/ Unavailable/ Low |
| **Technology:** Reference/Limited/Optimistic **CFE Zones:** Fine | 16 grid regions | Hourly via 120 representative hours | Reference/ High/ Low | Full/ Limited/ Full | Reference/ High/ Reference | Reference/ Unavailable/ Low |

*Table 1. Summary of scenario configurations modeled in this study. In this paper, Ref refers to the Reference Technology scenarios, Lim to the Limited Technology scenarios, and Opt to the Optimistic Technology scenarios.*

For CCS resources included in the Reference and Optimistic Technology scenarios, the capture rate of emissions (e.g., 90% or 95%) is used as the proportion of generation from these resources that qualify as CFE. As such, a company may or may not be able to report fully decarbonized scope 2 emissions if a generation resource using CCS is included in their mix of qualified CFE resources.

To compare between scenarios, system-level demand is kept constant. The analysis tested the impact that changes in system-level electricity prices might have on future loads and determined that expected price changes would lead to only small changes in modeled CFE procurement levels. This supports the choice to use equivalent system-level demand across the modeled scenarios.

[2] Even with no new CFE demand in the Counterfactual Reference scenario, note that state policies (i.e., Renewable Portfolio Standards [RPS] and Clean Energy Standards [CES]) drive some new CFE generation, sometimes referred to as EAC surplus [9].

Note that the impact of higher electricity prices is separate from analyzing directly the elasticity of EAC demand, which was not modeled directly in this study. In all scenarios, the level of CFE procurement is assumed to be inelastic to EAC prices (i.e., a fixed level of CFE procurement is modeled by region). However, the results of several of the scenarios suggest the potential for EAC prices to be substantially higher in future years than existing EAC spot prices. In a few instances, the resultant EAC prices are projected to be low or near zero. Future work could explore the potential impacts of elastic EAC demand as the range of EAC prices shown in this study could drive changes in CFE procurement.

Between each of the modeled deliverability zones, no electricity imports via interregional transmission are qualified to meet the modeled CFE demands. Within each of the modeled deliverability zones, this model assumes that transmission of CFE supply is feasible to any point in that zone. Thus, this study does not directly address the alternative proposal in GHGP's Scope 2 Consultation to pair attributes "with contracts or market instruments demonstrating physical delivery from the point of generation to the point of consumption," but it does give an upper bound on system conditions that could be expected should that alternative be chosen [3].

## 3. Results and Discussion

### 3.1. National Electricity Generation and Emissions Effects

This section focuses on the U.S. national results for the Reference Technology set of scenarios with new, incremental CFE procurement representing 10% of annual C&I demand in each 5-year increment between 2030 and 2050. Of interest are both the attributed and consequential impacts of CFE procurement. As discussed in Bistline et. al. [11], attributed generation refers to resources that can provide EACs to meet CFE requirements within a scenario. Consequential impacts are estimated as the difference between a CFE procurement scenario and a Counterfactual Reference scenario without procurement, representing the net power system impacts of adding CFE procurement. As such, consequential changes to generation and emissions can be estimated relative to a Counterfactual Reference scenario. Figure 2 shows snapshots of one of the Counterfactual Reference scenarios, with the generation and capacity mixes in each time period reflecting no incremental CFE procurement and Reference Technology assumptions.

Although there is no incremental CFE procurement included in the results of Figure 2, there is clear growth in CFE in meeting the projected national demand growth over the modeled years. This growth in CFE supply comes primarily from two sources: (i) new CFE capacity expected to be built under existing policies and programs (i.e., state-level RPS, CES, etc.); and (ii) new CFE generation expected to be built for economic reasons due to its cost effectiveness in the Counterfactual Reference scenario. The EACs underlying this expected new CFE generation in the future are not fully claimed and are geographically diverse, which drives the variations seen in the CFE procurement scenarios.

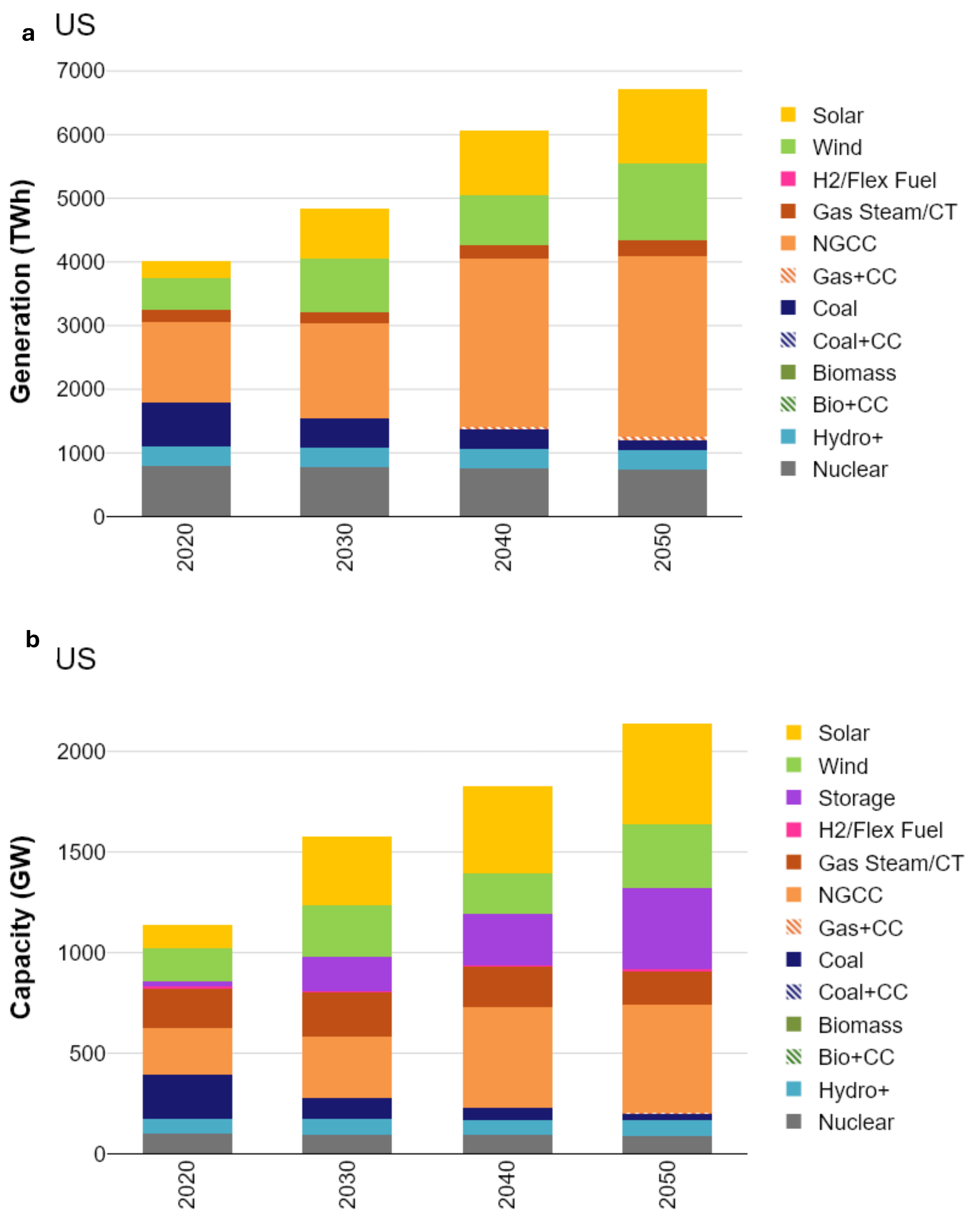

*Figure 2. Total U.S. electricity generation (top) and total capacity (bottom) projections in the Counterfactual Reference case with Reference Technology assumptions and no CFE procurement.*

Figure 3 illustrates changes relative to the Counterfactual Reference scenario in Figure 2 via results specific to the modeled year 2035 (results from all other modeled years with incremental CFE demand are directionally equivalent). Four distinct scenarios with different implementations of temporal and geographic deliverability matching are presented: annual matching with a national (most flexible) deliverability zone (i.e., CFE_GHGP1) reflecting the current GHGP Scope 2 Standard (2015), and hourly matching for three scenarios with increasingly narrow deliverability zones (CFE_Interconnect, CFE_Broad, and CFE_Fine).

For each scenario, the column with the subheading ‘A’ represents the amount and types of CFE generation that is attributed via EACs to the CFE procurement. Note that the total amount of attributed CFE generation is constant across the four scenarios. This is expected given that the design of this analysis maintains equivalent incremental CFE procurement levels in each scenario. EACs that are included in attributed generation are composed of three[3] blocks of CFE: (i) CFE deployed in the future in response to existing federal and state policies (e.g., state-level RPS, CES); (ii) CFE expected to be deployed due to its cost competitiveness compared to other generation resources; and (iii) new, incremental CFE capacity and generation deployed as a consequence of the GHGP’s proposed new spatiotemporal matching requirements. The last block of CFE is the only one that causes consequential changes to the integrated electricity system.

The column with the subheading ‘C’ represents the consequential changes in all generation sources when comparing each of these CFE procurement scenarios with the generation from the Counterfactual Reference scenario shown in Figure 2. The variation in the level of consequential changes across the scenarios ties directly to the availability of unclaimed EACs in the Counterfactual Reference scenario. Within all scenarios shown in Figure 3, generation from a mix of wind, solar, and natural gas with CCS is attributed to meeting CFE targets. In the most flexible scenario (i.e., annual matching combined with national deliverability in CFE_GHGP1), consequential generation and emissions impacts are more limited. While CFE procurement does incentivize additional natural gas with CCS, wind and solar resources are already competitive in several regions in the Counterfactual Reference scenario due to cost competitiveness and regional policy drivers. This provides a source of unclaimed EACs that count for attribution towards meeting CFE targets under this CFE program design and leads to consequential changes to power generation to be just 27% of the generation attributed to meeting CFE targets. The availability of unclaimed EACs decreases as the CFE program designs shift to increasing stringency of temporal matching and tighter geographic deliverability zones. As shown in Figure 3, this leads to consequential impacts becoming larger and more closely aligned with the generation attributed to meeting CFE targets. Nonetheless, for each scenario, the expected consequential changes to the power system are less than the expected attributional changes. Even in the CFE_Fine scenario, consequential changes to power generation are expected to be around 80% of the generation attributed to meeting CFE targets. This demonstrates how attributed changes to the power system associated with the use of EACs as a foundation for the MBM consistently overestimates more important consequential changes to the interconnected power system.

[3] Note that EACs associated with CFE that are claimed by entities pursuant to the GHGP proposed SSS and Legacy Contract “flexibility mechanisms” are excluded from these columns.

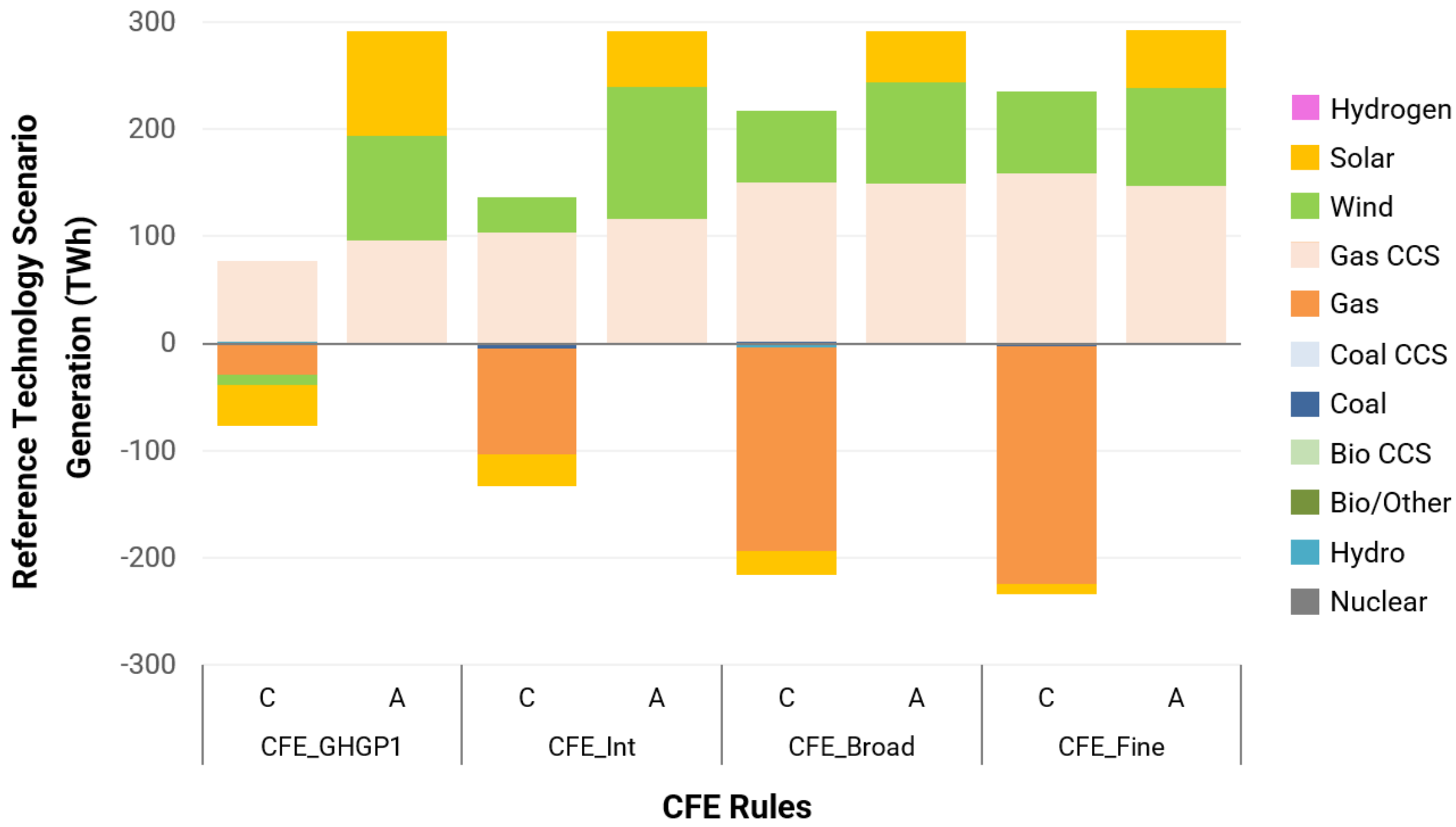

*Figure 3. Changes in national U.S. generation from CFE procurement in 2035 in the Reference Technology set of scenarios with new CFE procurement representing 10% of C&I demands. For each scenario, bar C represents consequential generation (change in generation relative to a Counterfactual Reference scenario with no CFE procurement), while bar A represents attributed generation (provides energy attribute certificates for qualification).*

Another important finding shown in Figure 3 is that a substantial portion of the consequential changes in generation deployed across Reference and Optimistic Technology scenarios can be expected to be met with deployment of natural gas with CCS. This result suggests that CCS may play an important role in providing clean firm capacity while supporting high levels of clean energy procurement and maintaining system reliability. However, there are meaningful uncertainties around technology readiness, the policy landscape for CCS, and the future emissions intensity that should be attributed to EACs associated with CCS-equipped generation. Given these uncertainties, CCS technologies were excluded from the Limited Technology scenarios to evaluate the ability of the power system to deliver CFE in the absence of CCS deployment, as discussed in a later section. This sensitivity analysis is not intended to imply that CCS will be unavailable or infeasible, but rather to assess the extent to which the modeled CFE pathways depend on CCS and to explore alternative technology portfolios that may emerge if CCS deployment is slower, more limited, or subject to different accounting and policy treatments than assumed in the Reference and Optimistic Technology scenarios.

### 3.2. Regional Effects of Deliverability-zone Definitions

Shifting focus from national to more granular regional results (as defined in Figure 1) reveals distinct impacts across individual regions, as shown in the maps of CFE procurement in Figure 4. For example, the South Atlantic region has the highest level of CFE procurement (37.6 TWh in 2035) across all the regions analyzed due to the expected strong growth in data center electricity demand. Under the most stringent deliverability scenario (i.e., Fine deliverability zones and hourly matching) shown in the upper left panel of Figure 4, the 37.6 TWh of CFE procurement in 2035 is met locally from electricity generated by qualifying CFE resources entirely within the South Atlantic region, primarily by natural gas with CCS resources in the Reference Technology scenario.

When the deliverability constraint is relaxed one level (i.e., Broad zone), expanding the deliverability zone to include electricity generated by qualifying CFE resources in the South Atlantic and two neighboring regions (i.e., Mid-Atlantic and Ohio Valley), the 37.6 TWh of CFE procurement in the South Atlantic region merges with CFE procurement from these two neighboring regions and is met collectively by CFE generation in the East region (see Figure 1 for the three levels of deliverability region boundaries). As shown in the upper right panel of Figure 4, in the Broad deliverability scenario, the Mid-Atlantic region supplies natural gas with CCS generation to CFE demand in the South Atlantic.

When deliverability is expanded further from the East region to the entire Eastern Interconnection as shown in Figure 1, multiple regions provide the generation needed to meet the collective CFE procurement target, as shown in the lower left quadrant of Figure 4. The CFE supply is led by wind generation in the MISO-North region, natural gas generation with CCS in the Mid-Atlantic and New York regions, and offshore wind generation in the New England region. Notably, the model incorporates state and regional policies driving clean energy deployment in these regions, including 13 GW of offshore wind capacity in New England[4] and an economy-wide net-zero $CO_2$ target in New York by 2050. Not all the generation from these state policies is claimed by existing demand for CFE, leaving some of the expected future clean energy generation available for the model to count as CFE procured to meet incremental CFE demand in the Eastern Interconnection in 2035 (and other modeled years). This is a good example of how attributed emissions reductions differ from consequential emissions reductions. For example, attributional emissions reductions include CFE generated as a result of existing policies that are unclaimed by CFE demand. These emissions reductions, however, would not count as consequential emissions reductions in this analysis because they would happen regardless of the new modeled increase in CFE demand. While the exact details of the generation resources that may be attributed to emissions reductions using the MBM may differ from this study when the real world reaches 2035, the concept holds that flexible deliverability zones lead to CFE deployment either (i) where resources are the least expensive to deploy and operate, such as natural gas with CCS being deployed near basins and geologic storage sites [18]; (ii) wind deployment in regions with high resource quality; or (iii) where qualifying CFE resources will be deployed in the future due to other drivers, such as state and regional emissions targets and clean energy standards.

[4] Although currently included in the model, it is unclear whether the state and regional policies driving offshore wind deployment will be retained. Future work could explore the impact to these results should this offshore wind capacity not be deployed.

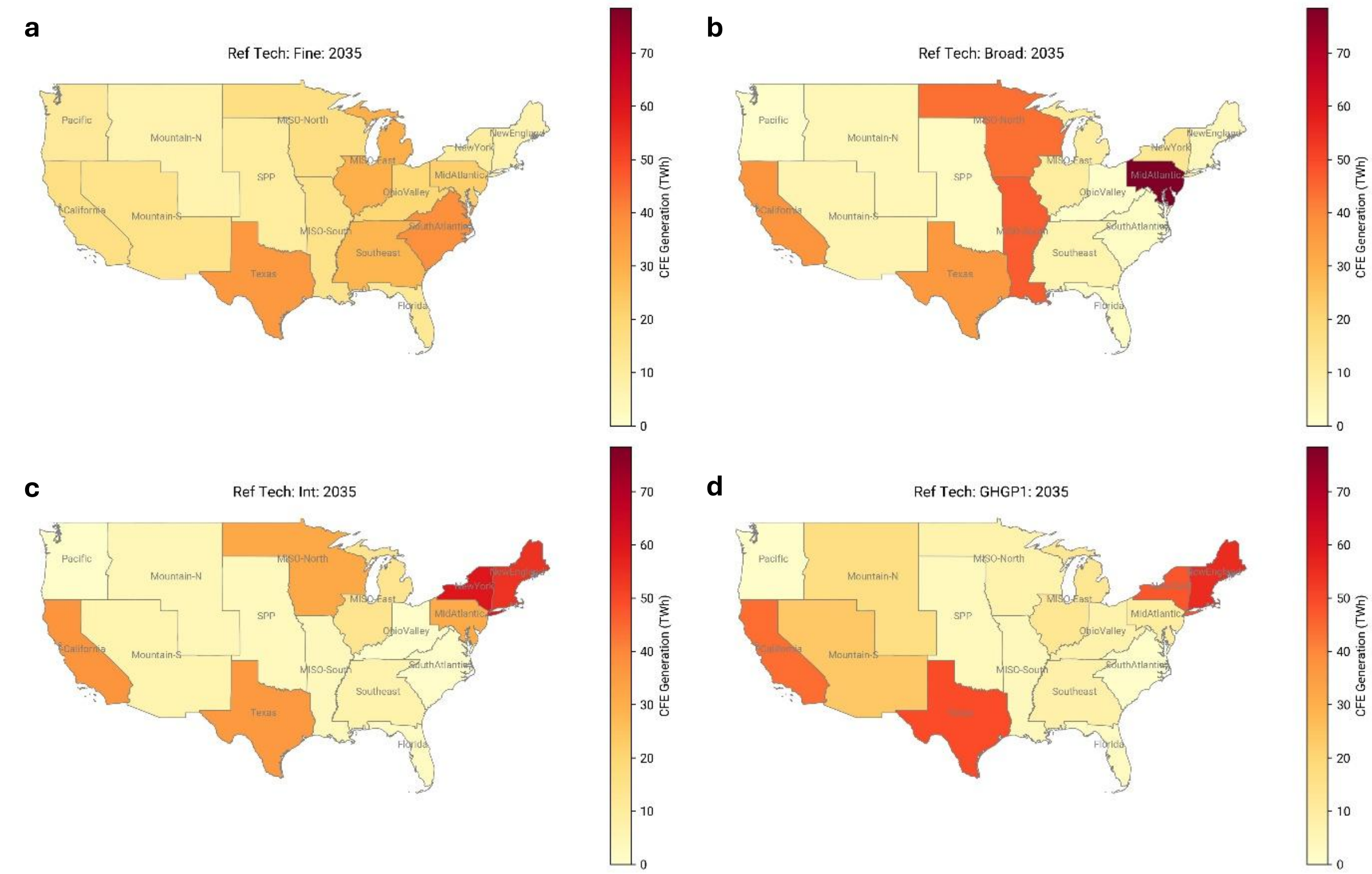

*Figure 4. Regional CFE generation in 2035 in the Reference Technology scenario set, across the four CFE Implementations.*

### 3.3. EAC Price Implications

Imputed EAC prices calculated in these scenarios reflect the incremental system costs with satisfying the CFE procurement constraint and can be interpreted as the marginal value of an additional EAC. In the annually matched, nationally deliverable scenario (GHGP1), generation-weighted EAC prices are low at $0.50/MWh in 2035 (see the leftmost column of each panel in Figure 5). As temporal matching and deliverability become more constrained (e.g., hourly vs annually; Fine deliverability zones vs larger geographic regions), generation-weighted EAC prices increase at the national level as shown in the upper left quadrant of Figure 5. This trend also is observed in other regions including the South Atlantic and Ohio Valley, as shown in the right half of Figure 5. However, this trend cannot be generalized because these results depend on tradeoffs (in costs, generation, and CFE procurement levels) between regions as the deliverability zones change. For instance, the Mid-Atlantic region in the lower left quadrant of Figure 5 shows *lower* EAC prices when the deliverability zone becomes *more constrained* going from the Broad to Fine definition in this study. This trend is caused by the lower relative costs of available generation technologies and the lower amount of C&I demand growth (and thus lower CFE procurement as defined in this study) as compared to the South Atlantic region.

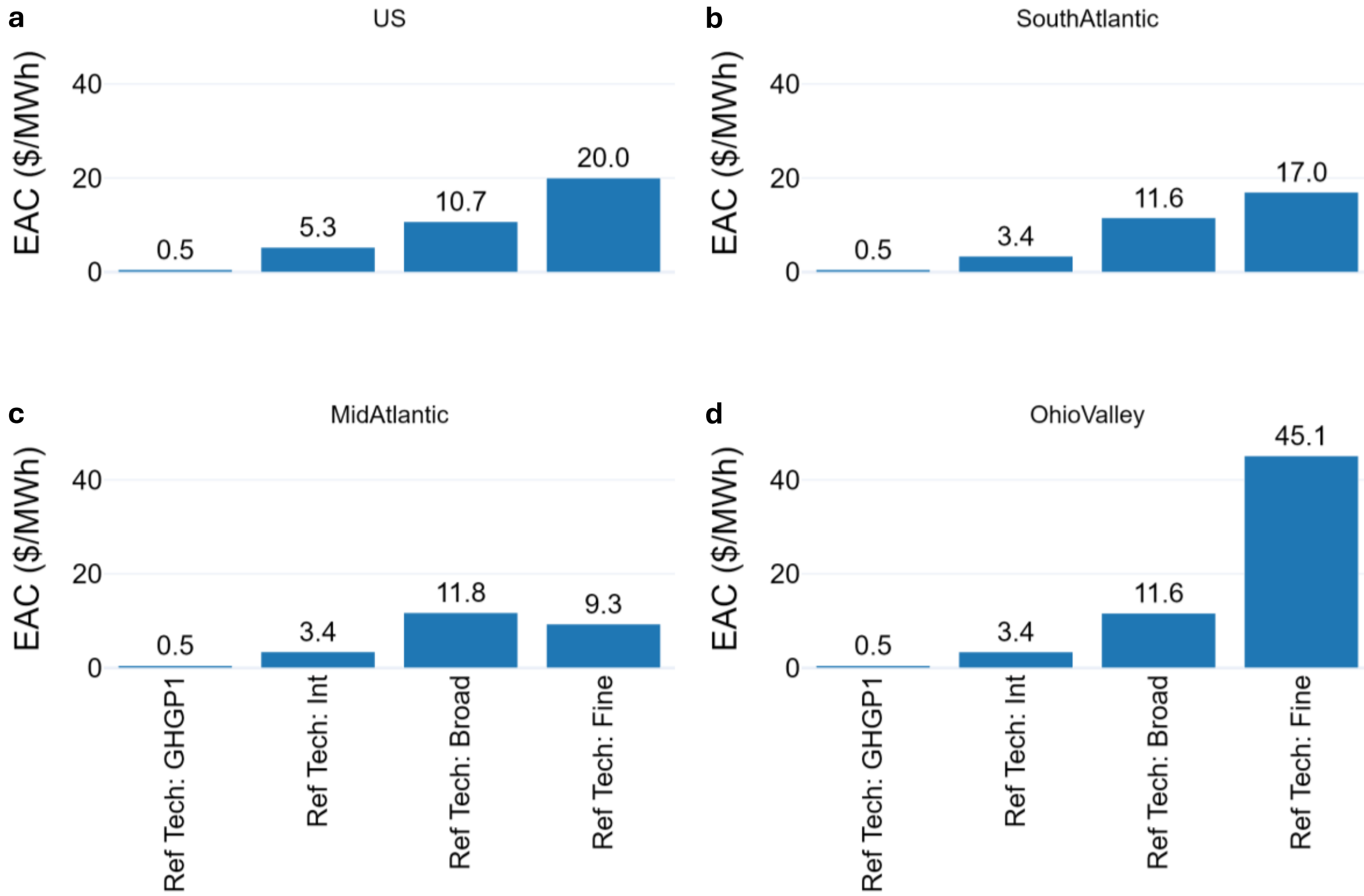

*Figure 5. EAC prices in 2035 across modeled scenarios with varying stringency of GHG protocol pillars for the U.S. (top-left), South Atlantic (top-right), Mid-Atlantic (bottom-left) and Ohio Valley (bottom-right) in the Reference Technology scenarios.*

### 3.4. Consequential Emissions Outcomes

The sourcing of CFE supply shown in Figure 4 reflects attributed generation that provides EACs to meet CFE requirements within a scenario. As discussed earlier, consequential impacts are estimated as the difference in system power generation resources and emissions between a CFE procurement scenario and a Counterfactual Reference scenario without the CFE procurement (though with equivalent total electric demand). The estimated change in consequential emissions in the Reference Technology set of scenarios is included in Table 2. When deliverability boundaries are tightened, less low and non-emitting incremental generation that would have been built in the Counterfactual Reference scenario is attributed to CFE procurement in the CFE procurement scenarios. This leads to an increase in consequential emissions reductions across the U.S. in these model results, reflected by moving from left to right along the bottom row of Table 2. As shown in Table 2, these modeling results suggest that the highest level of consequential emissions reductions occur in the scenario with hourly matching and Fine deliverability zones. This scenario also is the one with the highest anticipated EAC prices that could reduce future participation in CFE procurement.

| | Reference Tech: GHGP1 | Reference Tech: Interconnection | Reference Tech: Broad | Reference Tech: Fine |
|---|---|---|---|---|
| **South Atlantic Region** | (included in U.S.) | (included in Eastern Interconnection) | (included in East Region) | -8.2 |
| **East Region** | (included in U.S.) | (included in Eastern Interconnection) | -18.7 | -23.4 |
| **Eastern Interconnection** | (included in U.S.) | -24.7 | -52.5 | -59.3 |
| **U.S.** | -10.7 | -36.0 | -64.3 | -76.4 |

*Table 2. Changes in consequential emissions (Mt-$CO_2$/yr) for select regions in the Reference Technology scenarios. GHGP1 = National (current GHGP standard); Int = Interconnections; Broad = 6 ISO-level zones; Fine = 16 grid regions.*

Along with the decrease in consequential emissions reductions when deliverability zones increase in size is a decrease in the EAC prices. Comparing Figure 5 with Table 2, we see that broadening the deliverability regions from Fine to Broad reduces modeled consequential emissions reductions by 16%, while reducing the corresponding EAC price by nearly 50% at the national level. Similarly, relaxing the deliverability zone to the Interconnection level reduces the modeled consequential emissions reductions by 53% and reduces the EAC price by 74% compared to the scenario with Fine deliverability zones.

As previously mentioned, this study did not include an elasticity of EAC demand. At sufficiently low EAC prices like those calculated for the different conditions and locations in Figure 5, additional voluntary CFE procurement may arise. Alternatively, EAC prices perceived as too high may reduce voluntary CFE procurement.

### 3.5. Technology-sensitivity Results

To explore the robustness of these results, two additional technology sensitivities were analyzed: (i) a set of optimistic technology assumptions and (ii) a set of limited or conservative technology assumptions. The results under the Optimistic Technology conditions led to similar results as the Reference Technology conditions. The primary differences were lower EAC prices.

The results under the Limited Technology conditions led to more substantive differences between scenarios. For instance, by definition, natural gas with CCS is not available in the Limited Technology sensitivity, reflecting potential challenges either due to technical readiness or policy choices. This leads to changes in both the attributed and consequential generation as shown in Figure 6 compared to Figure 3. Primarily, wind and solar generation fills the full amounts of generation attributed to meeting CFE targets. The balance of consequential changes in the power system relative to the generation attributed to meeting CFE targets also changes in the Limited Technology scenario. For each deliverability zone shown, the consequential changes in generation as compared to the attributed changes are even smaller than similar changes shown in Figure 3. Regional CFE supply trends are similar in magnitude to Figure 4, with the underlying CFE supply attributed to meeting CFE targets mirroring the results in Figure 6.

Of note in Figure 6, the increased capital costs of natural gas capacity cause small amounts of additional coal generation to be dispatched by the model in the CFE procurement scenarios with Limited Technology conditions. Note that coal generation without CCS does not qualify as CFE in this study. Nonetheless, the calculation of consequential changes relative to the Counterfactual Reference scenario captures all generation changes that occur in the CFE procurement scenarios. This additional coal generation is cost-effective to maintain a balance of supply and demand of electricity across all represented hours in the Limited Technology scenarios.

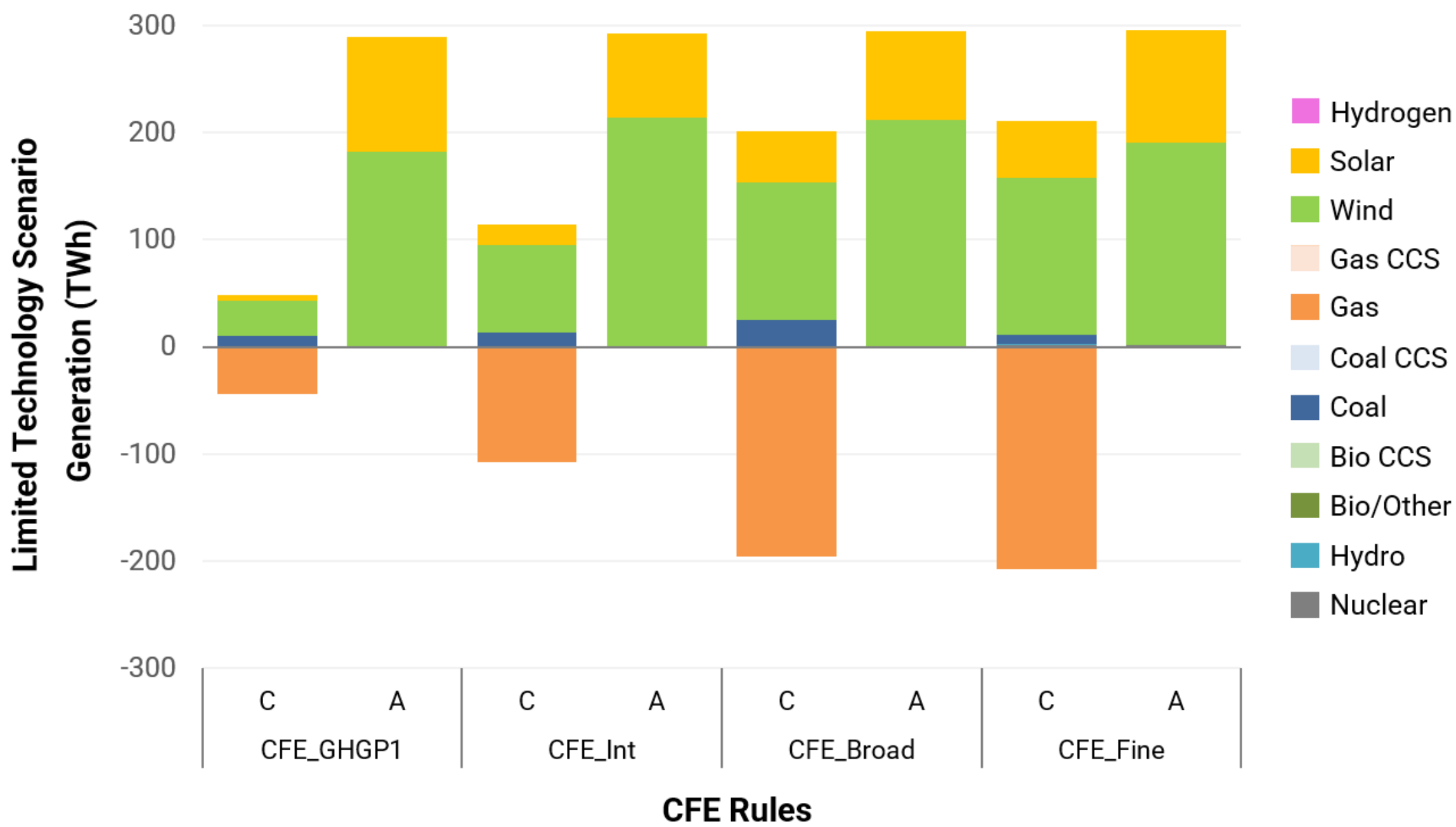

*Figure 6. Changes in national U.S. generation from CFE procurement in 2035 in the Limited Technology set of scenarios with new CFE procurement representing 10% of C&I demands. For each scenario, bar C represents consequential generation (change in generation relative to a Counterfactual Reference scenario with no CFE procurement), while bar A represents attributed generation (provides energy attribute certificates for qualification).*

As shown in Figure 7, EAC prices are noticeably higher in most cases under Limited Technology assumptions. Sharp increases in the South Atlantic EAC price in the most stringent deliverability case reflect the limited availability of wind and solar resources in this region. High EAC prices may drive down demand for voluntary CFE procurement in this region under these conditions, however elastic EAC demand was not modeled in this study. Consequential emissions reductions, shown in Table 3, are smaller in all cases as compared to those shown in Table 2. The Counterfactual Reference scenarios have similar emissions levels as a baseline: the Limited Technology scenario has 1,276 Mt-$CO_2$/yr as compared to 1,281 Mt-$CO_2$/yr for the Reference Technology scenario.

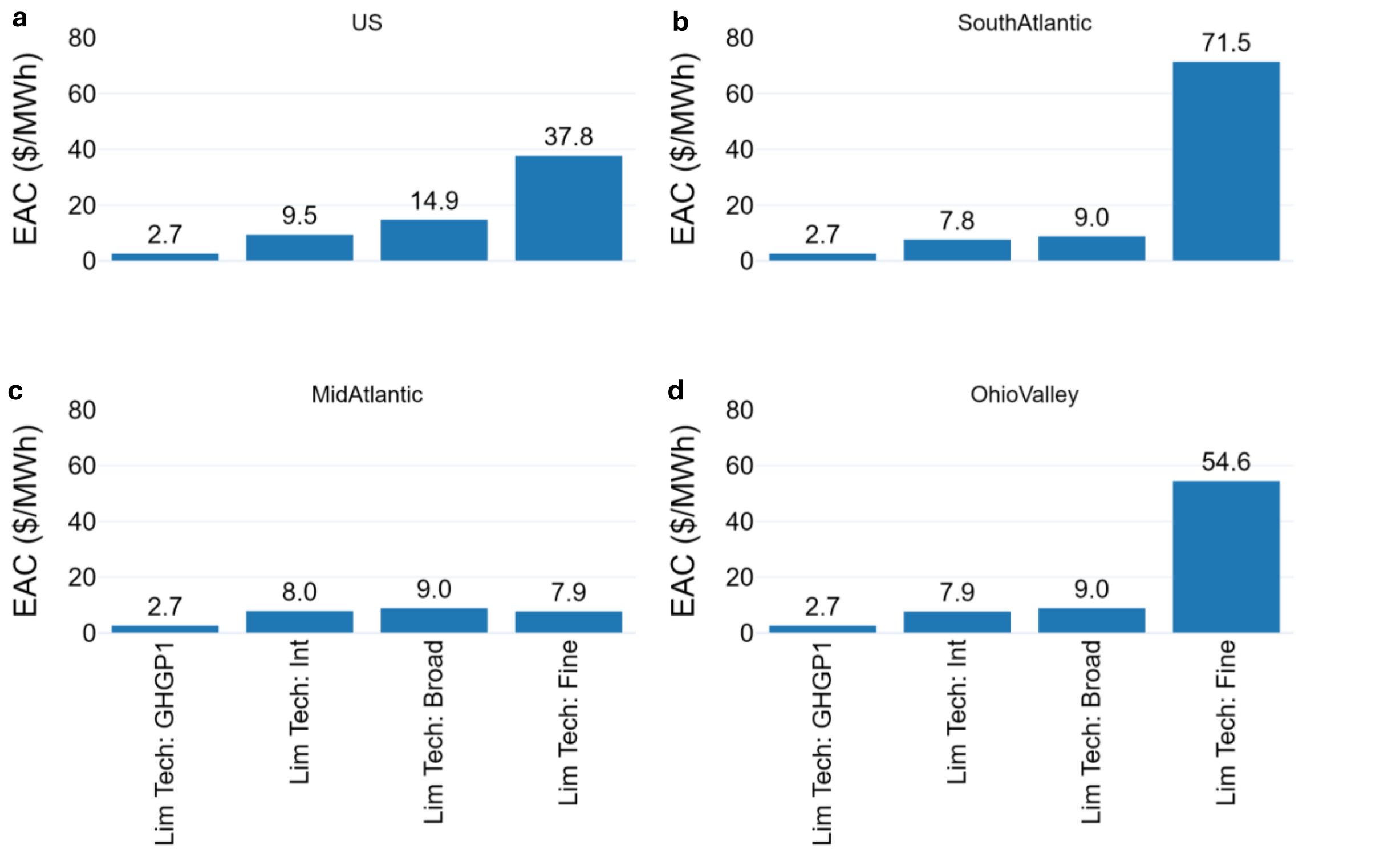

*Figure 7. EAC prices in 2035 across modeled scenarios with varying stringency of GHG protocol pillars for the U.S. (top-left), South Atlantic (top-right), Mid-Atlantic (bottom-left) and Ohio Valley (bottom-right) in the Limited Technology scenarios.*

| | Lim Tech: GHGP1 | Lim Tech: Int | Lim Tech: Broad | Lim Tech: Fine |
|---|---|---|---|---|
| **South Atlantic Region** | (included in U.S.) | (included in Eastern Interconnection) | (included in East Region) | -0.1 |
| **East Region** | (included in U.S.) | (included in Eastern Interconnection) | -9.8 | -16.8 |
| **Eastern Interconnection** | (included in U.S.) | -9.1 | -32.8 | -51.7 |
| **U.S.** | -6.1 | -23.7 | -48.3 | -69.1 |

*Table 3. Changes in consequential emissions (Mt-$CO_2$/yr) for select regions in the Limited Technology scenarios. GHGP1 = National (current GHGP standard); Int = Interconnections; Broad = 6 ISO-level zones; Fine = 16 grid regions.*

### 4. Conclusion and Policy Implications

The results suggest four policy-relevant implications for the design of Scope 2 market-based accounting rules and related clean electricity procurement programs.

First, geographic deliverability is a policy design lever, not only a technical accounting parameter. The choice of national, interconnection, broad regional, or fine grid-zone boundaries changes how much existing or policy-driven clean generation can be attributed to voluntary procurement and how much new consequential generation is induced by the accounting constraint.

Second, stricter deliverability improves alignment between attributed and consequential emissions outcomes but can increase EAC prices. This creates a participation tradeoff: rules that produce more physically grounded claims may also reduce voluntary procurement if buyers respond strongly to higher certificate or procurement costs.

Third, technology eligibility rules matter. The treatment of natural gas with CCS, nuclear, hydropower, biomass, storage, and other low- or non-emitting resources will influence both regional procurement costs and the resulting generation mix, so accounting standards may need to clarify not only where and when EACs can be matched but also which resources can provide qualifying attributes and under what emissions-rate assumptions.

Fourth, regional heterogeneity should be explicit in policy design. The modeled differences across the South Atlantic, Mid-Atlantic, Ohio Valley, East region, Eastern Interconnection, and the U.S. suggest that uniform deliverability rules may create uneven cost burdens and potentially affect future load-siting or procurement strategies.

Across most scenarios, geographically narrower deliverability zones are associated with greater consequential emissions reductions along with higher EAC prices. Under Reference Technology assumptions, expanding the spatial boundary from Fine to Broad reduces modeled consequential emissions reductions by 16% and reduces EAC prices nearly 50% at the national level. Further expansion of the spatial boundary to the Interconnect level reduces modeled consequential emissions reductions by 53%, with EAC prices falling to approximately 74% of those observed under the Fine spatial boundary. These results assume a fixed level of CFE demand, so further investigation of demand for CFE in response to changes in EAC prices could expand our understanding of this tradeoff of consequential emissions reductions and EAC prices.

National results show that in the Reference Technology scenario, the model preferentially deploys natural gas with CCS to meet new CFE demand (Figure 3). In the Limited Technology scenarios where natural gas with CCS is unavailable, wind and solar resources are deployed (Figure 6) with accompanying assumed higher capital costs (Table 1). Technology availability also impacts expected EAC prices, which at the national level range from $20/MWh in the Fine deliverability Reference Technology scenario to $38/MWh in the Fine deliverability Limited Technology scenario (Figure 5 & Figure 7). These results suggest restricted technology availability (or high technology costs) could reduce the consequential emissions reductions and raise EAC prices given the same level of CFE demand.

Individual regions show varying costs of $CO_2$ abatement under the scenarios, with the highest EAC prices occurring under Limited Technology assumptions and Fine spatial boundaries. This regional cost variation for CFE procurement could have implications for future load siting decisions or for which existing loads CFE will be procured to match.

Relaxing deliverability constraints in the South Atlantic region to allow electricity imports from adjacent regions as shown in Figure 1 reduces EAC prices and the sensitivity of EAC prices to technology assumptions. Under the Broad spatial boundaries, South Atlantic EAC prices converge with its neighbors in the East region to $9/MWh under Limited Technology assumptions (Figure 7) and $11.60/MWh under Reference Technology assumptions (Figure 5). Despite similar price outcomes, the underlying generation mix remains technology dependent: the Limited Technology scenario coupled with the Broad deliverability zone favors additional wind and solar expansion, while the Reference Technology scenario coupled with the Broad deliverability zone builds additional natural gas with CCS, in addition to smaller quantities of wind and solar. These findings highlight the importance of clearly defining “deliverability zones” in shaping both the emissions impact and cost of voluntary CFE procurement. They also suggest that system-level responses, including generation siting and technology choice, vary significantly across regions. This would have implications for total potential emissions reductions and system costs to achieve these reductions.

This study indicates that tighter temporal and geographic matching rules can increase consequential emissions reductions from voluntary clean electricity procurement, but these benefits are paired with higher modeled EAC prices and meaningful regional variation. The central implication is that Scope 2 accounting reforms will not only affect corporate reporting practice; they may also reshape clean electricity procurement markets, technology choices, regional investment patterns, and the credibility of emissions claims. Designing deliverability rules therefore requires balancing emissions integrity, procurement feasibility, market liquidity, and regional equity in implementation.

An important consideration not explicitly modeled in this study is the responsiveness of EAC demand to EAC price changes. If EAC demand is price-elastic, higher EAC prices could dampen CFE procurement, thereby reducing associated emissions reductions. Conversely, lower EAC prices may induce greater CFE procurement, amplifying system impacts. Incorporating EAC demand elasticity into future modeling could provide a more complete assessment of policy outcomes under the Scope 2 Consultation.

Additional research could also examine smaller subregional deliverability zones to better capture transmission constraints and congestion effects. Such refinements may further influence both modeled emissions and price formation.

**Funding**

This work was made possible through an unrestricted gift from Amazon. The authors have retained full editorial independence and are responsible for any errors or omissions.

**Credit Author Statement**

Daniel Livengood: Writing – original draft, Methodology, Investigation, Conceptualization. Arin Kaye*: Writing – original draft, Investigation, Conceptualization. Nils Johnson: Conceptualization, Methodology. John Bistline: Visualization, Methodology, Resources, Conceptualization. Aranya Venkatesh: Writing – original draft, Visualization. Bailie Neary: Writing – original draft, Visualization, Investigation, Conceptualization. Adam Diamant: Writing – original draft, Conceptualization.

**Supplementary Information**

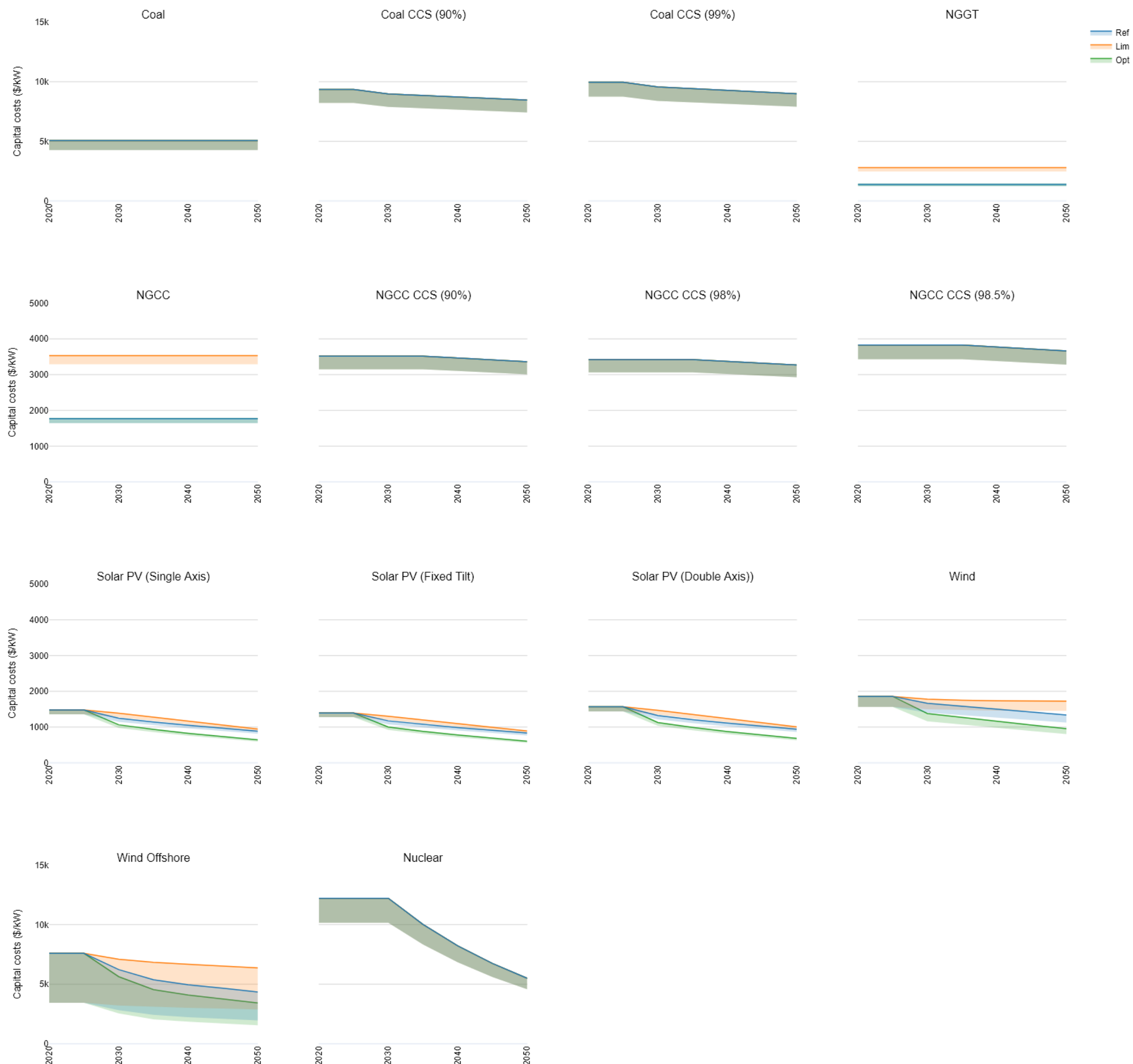

Figure S1 Capital cost assumptions for the Reference Technology scenarios (Ref), Limited Technology scenarios (Lim), and Optimistic Technology scenarios (Opt).

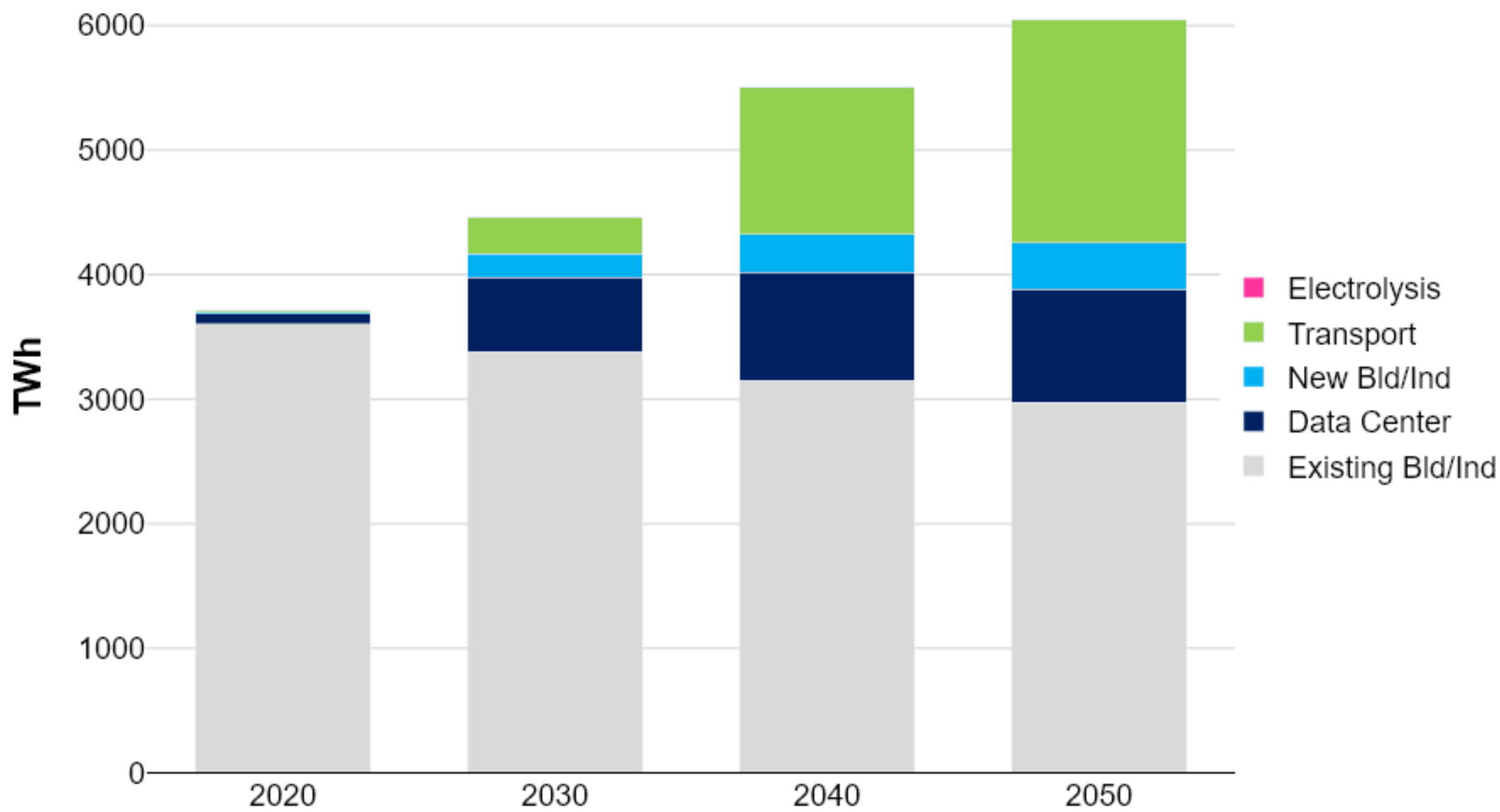

Figure S2 Total electricity demand for the U.S. broken out by sector. Bld refers to both residential and commercial buildings, and Ind refers to Industry.

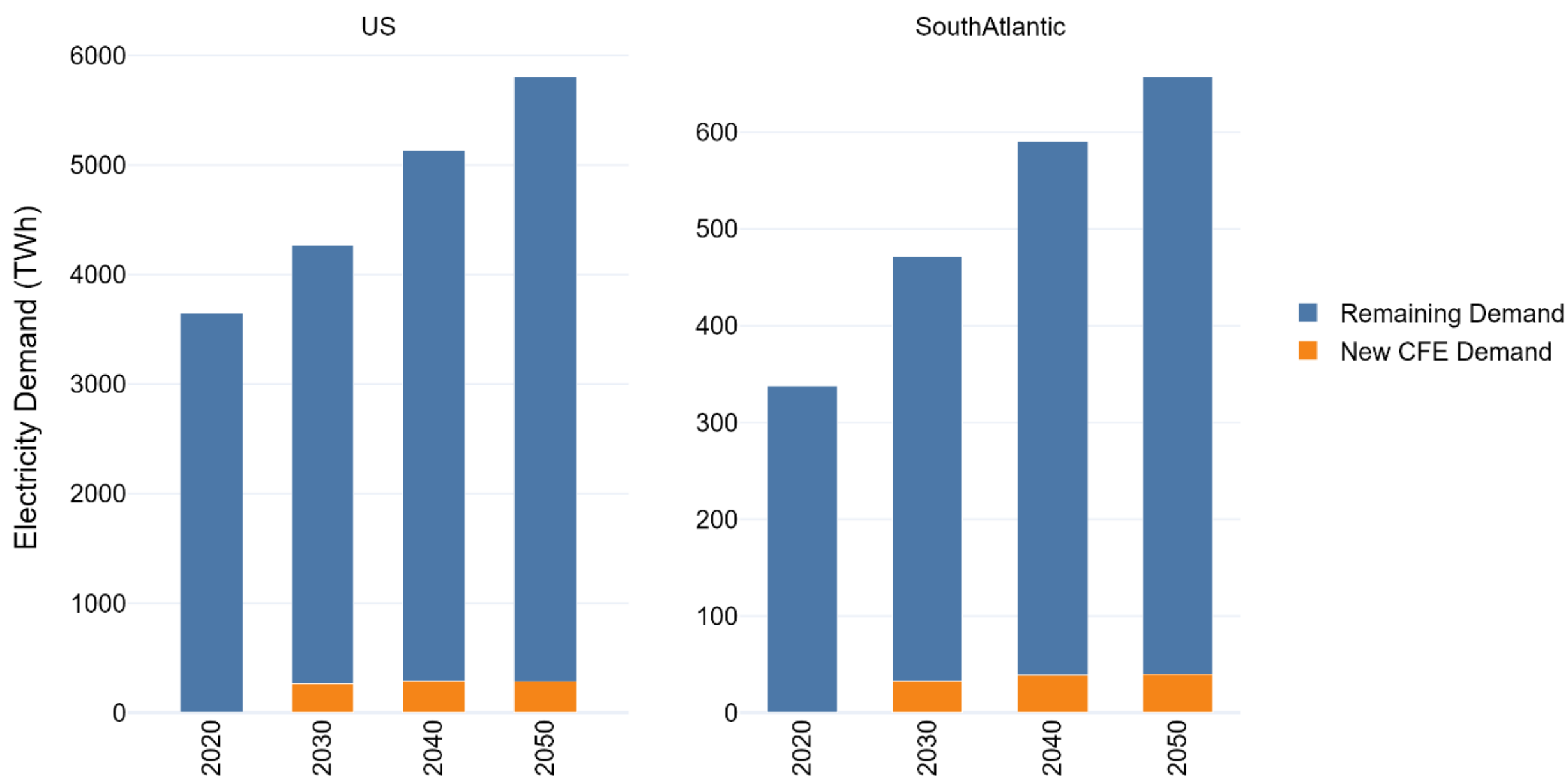

Figure S3: Total electricity demand and CFE procurement for the U.S. (left) and South Atlantic region (right) relative to total demand in each region in the modeled scenarios.

Comparing Figure 2 and Figure S3, there is a noticeable gap between the total amount of electricity generation and demand in each modeled year. This is caused by a combination of transmission and distribution losses from the point of generation to the point of consumption along with, in later modeled years, an increase in losses caused by time-shifting energy via energy storage resources. Lithium-ion batteries, for instance, have a roundtrip efficiency of around 85%, meaning that for every 1 MWh of generation used to charge those batteries, approximately 0.85 MWh will be available for discharge later due to the electrochemical conversion processes of the batteries.

As shown in Figure S3, the decision to model incremental CFE procurement as 10% of C&I demand led to annual, national, incremental CFE demand remaining relatively flat across the modeled years of 2030 to 2050 based on the underlying demand assumptions in US-REGEN shown in Figure S2. This is caused by modeled efficiencies and other demand-side changes in the commercial and industrial sectors offsetting most of the anticipated demand growth from data centers. The sector driving most of the demand growth shown in Figure S3 is from the electrification of the transportation sector. Variations in the potential progression of incremental CFE demand over time could be explored in future work.